# Overload Control in SIP Networks: A Heuristic Approach based on Mathematical Optimization

Ahmadreza Montazerolghaem[1], Mohammad Hossein Yaghmaee Moghaddam[1], Farzad Tashtarian[2]

[1]Department of Computer Engineering, Ferdowsi University of Mashhad, Mashhad, Iran

[2]Young Researchers and Elite Club, Mashhad Branch, Islamic Azad University, Mashhad, Iran

Ahmadreza.montazerolghaem@stu.um.ac.ir, hyaghmae@ferdowsi.um.ac.ir, f.tashtarian@mshdiau.ac.ir

*Abstract*— The Session Initiation Protocol (SIP) is an application-layer control protocol for creating, modifying and terminating multimedia sessions. An open issue is the control of overload that occurs when a SIP server lacks sufficient CPU and memory resources to process all messages. We prove that the problem of overload control in SIP network with a set of *n* servers and limited resources is in the form of NP-hard. This paper proposes a Load-Balanced Call Admission Controller (LB-CAC) based on a heuristic mathematical model to determine an optimal resource allocation in such a way that maximizes call admission rates regarding the limited resources of the SIP servers. In fact, LB-CAC by having some critical information of SIP servers determines the optimal "call admission rates" and "signaling paths" for admitted calls along optimal allocation of CPU and memory resources of the SIP servers through a new linear programming model. A comparison between the numerical and experimental results implies the efficiency of the proposed method.

## I. INTRODUCTION

SIP is an application layer signaling protocol standardized by IETF for initiating, modifying, and tearing down multimedia sessions [1]. The architecture of SIP protocol consists of two entities including User Agent (UA) and SIP Server (briefly referred as server). UAs themselves are divided into User Agent Client (UAC) and User Agent Server (UAS) that transmit signaling messages of request and response, respectively. A request message along with all its related responses is called as SIP transaction. SIP servers are designed in such a way that both record (stateful) or do not record (stateless) the history of transaction operations for every request. The messages and their exchange process for initiating and tearing down a session are illustrated in Fig. 1. Call setup message or "INVITE" (briefly referred as Call Request) is the most important request message since its transaction has maximum process load on CPU server [7, 9]. Once a session is created, media such as audio and video are exchanged between the parties without passing from the servers, so that the only load on the server would be signaling messages.

Considering the widespread use of this protocol and its numerous (several millions) users in the near future, it is necessary to assess the performance of SIP servers in overload states. SIP overload occurs whenever SIP server lacks enough resources to process all messages [2, 3]. In this regard, the most important resources are processing power of CPU and memory [4].

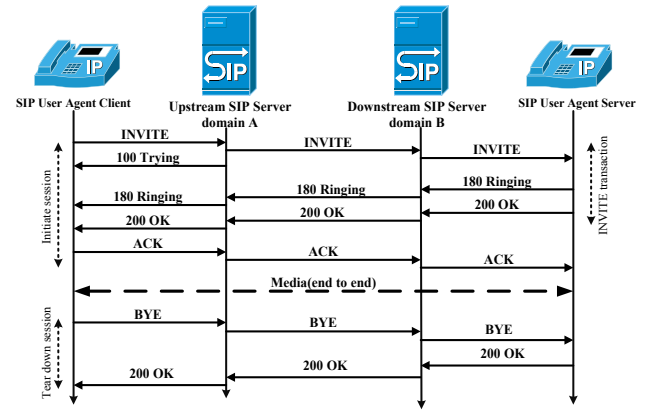

Fig. 1. Transmission of SIP signaling messages for initiating and tearing down a session by using the upstream and downstream SIP servers

### A. Processing power of CPU and memory of SIP server

Each SIP server has a limited capacity and therefore, its total capacity decreases when an overload occurs. This is because most of its processing resources are allocated for rejecting or processing the messages that would be ultimately rejected [5]. SIP servers make use of their reliable mechanism whenever, in particular, they are applied to an unreliable transmission protocol such as UDP [4]. To this end, it needs to provide reliability by retransmission of the messages with unconfirmed delivery [5]. In this state, a large set of retransmission timers are employed [2, 4]. Although this mechanism is useful in case of unreliable links, under the overload conditions, it imposes high loads to the server and decreases its efficiency [4][1]. This is because the redundant retransmissions and manipulation of the mentioned timers increase CPU and memory occupation and worsen server overload. In this state, the server's queue would be filled with duplicate messages and the processing power of the server would be spent on processing and transmitting duplicate messages or answering the expired timers. On the other hand, the base mechanism of SIP protocol which lacks required efficiency for overload control would be activated [2]. In this method, once the server reaches its maximum capacity, it rejects new call request messages by issuing "503 service unavailable" message. The cost of this method is not negligible as compared with giving service to a request. In addition, for servers configured statefully (as the dominant configuration), some state information are stored for each transaction. In case of no supervision over the number of established calls, the entire memory of the server might be engaged and no memory is remained to allocate for the new calls.

---

[1] Nevertheless, UDP is more common, as compared to TCP, for transmission of SIP messages. Because TCP suffers from high delay and scalability issues which are not acceptable for the real-time signaling protocols [1, 2].

As previously discussed, both saturation of processors and memory shortage in overload state may degrade the performance of SIP servers. Therefore, to obtain maximum capacity and prevent overload occurrence, it is necessary to make optimum use of resources and prevent its waste in tasks such as rejecting or responding messages that would be ultimately rejected. The main goal pursued in this work is how to optimally allocate SIP servers' resources to the admitted calls.

In section 2, the recent work studying overload control in SIP networks are pointed out. System model and problem formulations (section 3), presentation of the proposed method (section 4), performance evaluation (section 5), and conclusion and future work (section 6) comprise the next parts of this study.

## II. RELATED WORK

One approach to overload avoidance is the distribution of new input traffic over the SIP servers based on their accessible capacity by using a "Load Balancer" [6-8]. Taking this approach, consequently, reduces the probability of overload occurrence in a given server. However, the performance of these methods depends on Load Balancer's efficiency because the whole signaling traffic of SIP network passes through it. As a result, Load Balancer itself is threatened by the risk of overload in case of severe overloads and therefore its capacity needs to be enhanced by the available techniques.

Another approach for dealing with overload issue is to apply overload control methods, which are divided into two categories: local and distributed. Through the local overload control methods, the overloaded server itself has a consistent control over its resources usage without any need for interaction with other network servers and applies its controlling method, and rejects the excess calls. The criteria for identifying the overload in some of these methods are queue length and CPU usage level. According to these criteria, a set of thresholds are defined, exceeding which makes the server to enter overload stage, so that it starts to reject the incoming calls. The main drawback of this method is that the cost of call reject cannot be ignored, and when dealing with heavy overloads, the server must use its entire resources for rejecting the excess calls. Accordingly, the server will not be able to respond any services [9]. On the other hand, depending that the overloaded server whether notifies its status to the upstream servers or the upstream servers are themselves informed of the current status of the downstream (overloaded) server, the distributed methods are categorized into two explicit and implicit methods [3]. Based on the information exchanged among the servers, these methods are also classified into the rate-based, loss-based, signal-based, window-based, and on/off control techniques [11]. Within the rate-based techniques, downstream server has a consistent supervision over the rate of messages delivered from total upstream servers and announces its admissible rate to them [5, 10]. In loss-based technique, downstream server frequently measures its current load and accordingly requests the upstream servers to reduce their transmitted load [12]. In window-based methods, the load is not transmitted to the downstream server, unless there are empty slots in upstream server window. The main issue of these methods is window size which is adjusted using the feedback of the downstream server [1, 2, and 13]. The signal-based methods are those in which the upstream server reduces its transmission rate when receiving "503 Service Unavailable" message in order to prevent further transmission of 503 message from the downstream server [14]. Unlike signal-based method which is based on not using "Retry-After" header, within the on/off control method a given server can either hold off or on its received load by transmission of Retry-After feedback. However, the hold-on or hold-off state of the received traffic flow can be also controlled using the other mentioned methods [15].

The third approach applied for overload control is based on retransmission rate methods which review retransmission mechanism of SIP by studying servers' buffer size [4, 16]. By limiting the dedicated memory of the server, it can be prevented from admitting the over-capacity calls. However, this policy loses its efficiency once the call rate rises, as the server processor is forced to analyze the messages to recognize their content. Therefore, under such conditions, the server reaches saturation, which typically occurs under the higher loads.

The next approach to overload avoidance is based on TCP flow control for the purpose of regulating SIP overload [17]. However, problems such as scalability and high delay hinder use of this protocol. Furthermore, the congestion designed for TCP control mechanism occurs due to limited bandwidth, whereas the overload in SIP is caused by the limited processing capacity of servers' CPU [4].

Based on the mentioned points, it can be stated that the main disadvantages of the present overload control approaches are: Firstly, their reliance merely over the local call reject reduces their throughput. Secondly, for the majority of explicit feedback-based methods, continuous revision of the status and feedback calculation (which is the function of overloaded server) has complexity and header. The third defect of these methods is the delay in feedback arrival to the upstream, which results in instability of these methods. Nevertheless, these methods are more accurate in overload detection as compared to the implicit methods. Therefore, overload detection, feedback generation, and running the overload control algorithm incurs CPU and memory usage costs and affects throughput of the server. The contributions of this work are summarized as follows:

(1) Proving the NP-hard nature of the overload control problem for $n$ number of servers with limited memory and CPU; (2) Presenting a Load-Balanced Call Admission Controller (LB-CAC) for optimum allocation of server resources; (3) Proposing a mathematical model to maximize the performance of SIP severs in terms of throughput and the usage of their resources; and (4) Implementing the proposed method in a testbed and comparing the experimental and simulated results.

## III. SYSTEM MODEL AND PROBLEM FORMULATION

In general, SIP network consists of a set of $n$ SIP servers with limited processing and memory resources. Each server uses its resources to make session between the local users and users in other domains. In this paper, it is assumed that the binary symmetric matrix $L_{ij}, i,j \in \{1,...,n\}$ makes the topology of SIP server network. In this matrix, $L_{ij} = 1$ implies presence of a physical link between servers $i$ and $j$, while $L_{ij} = 0$ indicates that the two given SIP servers are not adjacent through a direct link. Note that the main diagonal of this matrix is zero. Assume a two-dimensional array $\mathbb{C}$ with the size of $(n \times n)$, where $\mathbb{C}^{ii}$ and $\mathbb{C}^{ij}$ demonstrate the number of local calls in server $i$ (where both caller and callee are registered in a same server) and the number of outbound calls from server $i$ to server $j$, respectively. Let $C^{ij}$ be the optimal number of admitted calls established from server $i$ to server $j$ where $C^{ij} \leq \mathbb{C}^{ij}$. Regarding the optimal value of $C^{ij}$, $R^{ij}_{kl}$ shows the number of calls from origin $i$ to destination $j$ which must be relied from server $k$ to server $l$ (see Fig.2).

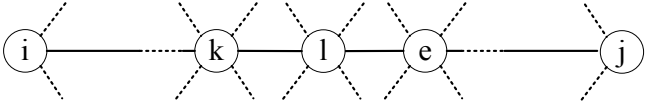

Fig. 2. Transmitting calls from the origin $i$ to the destination $j$ assisted by sip servers $k$ and $l$

To perform its operation, each server $i$ relies on its remaining CPU and memory resources, which are denoted as $P_i$ and $M_i$, respectively.

The duty cycle of Load-Balanced Call Admission Controller (LB-CAC) is presented in Fig. 3. At the beginning of each time slot $\tau$, all servers transmit the amount of remaining resources and the number of new local and outbound requests to LB-CAC through the UDP packets. This critical information are gathered in $t_g$ units of time. By proposing a mathematical linear programming model in the next part, LB-CAC determines the optimal values of $C^{ij}$ and $R^{ij}_{kl}$ and then broadcasts the results to the servers in $t_c$ and $t_n$, respectively (Fig. 3). After that, LB-CAC enters into the idle state and waits $t_{idle}$ units of time for the next time slot $\tau$. In fact, a new incoming call during $\tau$ should be entered into hold-on state by SIP server until the next $\tau$. Compared to a duration of a call, this time is negligible which can also be shortened by reducing the $\tau$ time. Regarding the definition of the duty cycle of LB-CAC, the number of calls per $\tau$ is called flow.

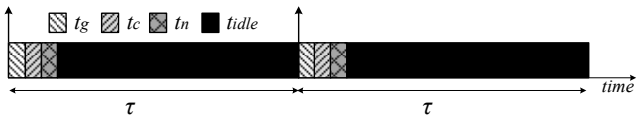

Fig. 3. The duty cycle of LB-CAC

## IV. LB-CAC

In this section, LB-CAC controller is designed in a way to optimally determine $C^{ij}$ and $R^{ij}_{kl}$ based on a new mathematical linear optimization model. Before proposing the model, it would be proved that the problem of overload controlling in SIP networks is NP-hard.

*Theory 1*: Overload control problem for a set of $n$ SIP servers with limited CPU and memory resources is in the form of mixed integer nonlinear programming (MINLP) model, which is generally NP-hard ($n > 2$).

*Proof*: Let $B^{ij}_{kl}$ be a binary variable in which $B^{ij}_{kl} = 1$ states that a call request from server $i$ to server $j$ can be transmitted from server $k$ to server $l$, unless ($B^{ij}_{kl} = 0$) server $k$ and $l$ do not participate in transferring request from server $i$ to server $j$. Regarding the limited hardware resource in the servers, the optimal values of $B^{ij}_{kl}, C^{ij}$, and $R^{ij}_{kl}$ can be obtained through a following MINLP model:

$$\max \sum_{i=1}^{n} \sum_{j=1}^{n} C^{ij} \quad (1)$$

Subject to:

$$C^{ij} \leq \mathbb{C}^{ij}, \quad \forall i,j \quad (I)$$
$$\sum_{k=1}^{n} B^{ij}_{kl} R^{ij}_{kl} = \sum_{e=1}^{n} B^{ij}_{le} R^{ij}_{le}, \quad \forall i,j,l, i \neq l, j \neq l \quad (II)$$
$$\sum_{k=1}^{n} B^{il}_{kl} R^{il}_{kl} = C^{il}, \quad \forall i,j,l, i \neq l \quad (III)$$
$$\sum_{e=1}^{n} B^{lj}_{le} R^{lj}_{le} = C^{lj}, \quad \forall i,l,j, j \neq l \quad (IV)$$
$$R^{ii}_{kl} = 0, \quad \forall i,k,l \quad (V)$$
$$R^{ij}_{ki} = 0, \quad \forall i,j,k \quad (VI)$$
$$B^{ij}_{kl} - L_{kl} \leq 0, \quad \forall i,j,k \quad (VII)$$
$$\alpha_1 C^{ll} + \alpha_2 (\sum_{i=1}^{n} \sum_{j=1}^{n} \sum_{k=1}^{n} B^{ij}_{lk} R^{ij}_{lk} + \sum_{i=1}^{n} \sum_{j=1}^{n} \sum_{k=1}^{n} B^{ij}_{kl} R^{ij}_{kl}) \leq P_l, \quad \forall l \quad (VIII)$$
$$\beta_1 C^{ll} + \beta_2 (\sum_{i=1}^{n} \sum_{j=1}^{n} \sum_{k=1}^{n} B^{ij}_{lk} R^{ij}_{lk} + \sum_{i=1}^{n} \sum_{j=1}^{n} \sum_{k=1}^{n} B^{ij}_{kl} R^{ij}_{kl}) \leq M_l, \quad \forall l \quad (IX)$$

Variables: $B^{ij}_{kl} \in \{0,1\}, C^{ij}, R^{ij}_{kl}, P_l, M_l \geq 0, \quad \forall k,l,i,j$

Constraint (I) limits the number of admitted calls to the number of existing call requests. Constraint (II) makes a trade-off between the input and output flows; meaning that the total input and output flows in server $l$ must be equaled for a couple of origin and destination. Constraint (III) aggregates the total input flows to server $l$ from origin $i$ passing through its neighbors. The next constraint, distributes the total output flows from server $l$ to server $j$ among its neighbors. Constraint (V) prevents flow with the same origin and destination. Constraint (VI), which restricts creation of some loops in a given path, will be discussed later on. Constrain (VII) limits the binary $B^{ij}_{kl}$ variable in such a way that if $L_{kl}$ equals zero, $B^{ij}_{kl}$ must be set to zero, unless if $L_{kl}=1$, $B^{ij}_{kl}$ can be either zero or one. The next two constraints consider the limited SIP servers' resources: Constraint (VIII) allocates the residual processing power of server $l$ with coefficients $\alpha_1$ and $\alpha_2$ to establish local and outbound calls of server $l$; similarly, constrain (IX) allocates residual memory of server $l$ with coefficients $\beta_1$ and $\beta_2$ to establish local and outbound calls of server $l$. Note that two variables $R^{ij}_{lk}$ and $R^{ij}_{kl}$ have an identical influence on servers' resources. To estimate parameters $\alpha_1, \alpha_2, \beta_1$, and $\beta_2$, two linear programming models will be proposed in the next part.

Although the objective function and constraints I, V, VI, and VII are linear, presence of nonlinear constraints II, III, IV, VIII, and IX and the binary $B^{ij}_{kl}$ variables has made the problem in MINLP model which is NP-hard [18, 19].

## A. Proposed heuristic method

Before elaborating the proposed method, first consider Fig. 4 with origin server 1 and destination server 3. In this study, two types of loop are presented between the origin and the destination: k-hop source loop (SL) and k-hop non-source loop (nSL). In the former type, the origin server participates in the loop, while in the later, the loop is created by some servers expect the origin server. The proposed model in Eq. (1) can only restrict SLs, as it applies constraints (V) and (VI).

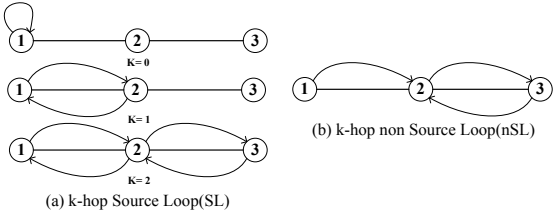

Fig. 4. Sl and nSL loops between the origin and destination flow path

Although the proposed model in Eq.(1) increases the total number of admitted calls, it can be easily shown that considering $\sum_{i=1}^{n}\sum_{j=1}^{n} C^{ij}$ as the objective function, it is not possible to prevent nSLs. Furthermore, as previously discussed, this nonlinear model is NP-hard and is unsolvable in polynomial time [19]. In this regard, the objective function should be modified in such a way that not only it maximizes the total number of admitted calls, but also it minimizes CPU and memory usages. Since the loop creation requires resource usage, the next model should prevent nSL. As a result, by modifying the objective function and some constraints, we can convert the proposed model in Eq. (1) into a linear programming (LP) model as follows:

$$max \; \gamma \frac{\sum_{i=1}^{n}\sum_{j=1}^{n} C^{ij}}{\sum_{i=1}^{n}\sum_{j=1}^{n} \mathbb{C}^{ij}} - \varphi\left(\frac{\sum_{l=1}^{n}(p_l)}{\sum_{l=1}^{n}(P_l)} + \frac{\sum_{l=1}^{n}(m_l)}{\sum_{l=1}^{n}(M_l)}\right) \quad (2)$$

Subject to:
$$C^{ij} \leq \mathbb{C}^{ij}, \qquad \forall i,j \; (I)$$
$$\sum_{k=1}^{n} L_{kl} R_{kl}^{ij} = \sum_{e=1}^{n} L_{le} R_{le}^{ij}, \qquad \forall i,j,l, i \neq l, j \neq l \; (II)$$
$$\sum_{k=1}^{n} L_{kl} R_{kl}^{il} = C^{il}, \qquad \forall i,l, i \neq l \; (III)$$
$$\sum_{e=1}^{n} L_{le} R_{le}^{lj} = C^{lj}, \qquad \forall l,j, j \neq l \; (IV)$$
$$R_{kl}^{ii} = 0, \qquad \forall i,k,l \; (V)$$
$$R_{ki}^{ij} = 0, \qquad \forall i,j,k \; (VI)$$
$$\alpha_1 C^{ll} + \alpha_2 (\sum_{i=1}^{n}\sum_{j=1}^{n}\sum_{k=1}^{n} L_{lk} R_{lk}^{ij} +$$
$$\sum_{i=1}^{n}\sum_{j=1}^{n}\sum_{k=1}^{n} L_{kl} R_{kl}^{ij}) \leq p_l, \qquad \forall l \; (VII)$$
$$\beta_1 C^{ll} + \beta_2 (\sum_{i=1}^{n}\sum_{j=1}^{n}\sum_{k=1}^{n} L_{lk} R_{lk}^{ij} +$$
$$\sum_{i=1}^{n}\sum_{j=1}^{n}\sum_{k=1}^{n} L_{kl} R_{kl}^{ij}) \leq m_l, \qquad \forall l \; (VIII)$$
$$p_l \leq P_l, \qquad \forall l \; (IX)$$
$$m_l \leq M_l, \qquad \forall l \; (X)$$
Variables: $C^{ij}, R_{kl}^{ij}, p_l, m_l \geq 0, \quad \forall i,j,k,l$

Connectivity between two servers is developed using the matrix $L_{ij}$. The loose upper bound of the optimal CPU and memory usages are denoted with $p_l$ and $m_l$, respectively; where, based on constraints (IX) and (X), they can reach to the maximum values of $P_l$ and $M_l$ as the strict power bound. Constraints (VII) and (VIII) will also help the objective function to prevent creation of extra $R_{kl}^{ij}$, SL and nSL loops. Coefficients $\gamma$ and $\varphi$ indicate the significance of variables $C^{ij}$ and resource usage ($p_l$ and $m_l$) in the objective function, respectively. In other words, they make a trade-off between the resources usage and network throughput. Ultimately, this model enhances the total throughput of the servers, without any overload, using the optimal resource allocation. Here, the overall throughput can be considered as the total flow passing through the servers. Therefore, considering the servers' resources, LB-CAC tries to maximize call admission through distributing calls among the servers.

## B. Calculation of α and β

The two following linear models are designed to approximate the coefficients $\alpha$ and $\beta$. These coefficients indicate the impact of establishing the local and outbound calls of a given server regarding its resource usage. By considering a dataset consists of $h$-tuple of $(\bar{C}_q^{ii}, \bar{R}_{ij,q}^{ij}, \bar{p}_q, \bar{m}_q)_{q=1,...,h}$, the proposed model determine the optimal values of $\alpha$ and $\beta$, where $\bar{C}_q^{ii}, \bar{R}_{ij,q}^{ij}, \bar{p}_q$, and $\bar{m}_q$ indicating the number of local calls in server $i$, number of outbound calls of server $i$, CPU and memory usages in server $i$, respectively. By considering the first constraint, the objective functions in these two models are designed in a way that minimizes the sum of difference between measured values of $\bar{p}_q$ and $\bar{m}_q$ as $\alpha_1 \bar{C}_q^{ii} + \alpha_2 \bar{R}_{ij,q}^{ij}$ and $\beta_1 \bar{C}_q^{ii} + \beta_2 \bar{R}_{ij,q}^{ij}$, respectively. Here, constraint (II) normalizes constraint (I).

$$min \; \sum_{q=1}^{n} x_q \qquad (3)$$
Subject to:
$$\bar{p}_q - \left(\alpha_1 \bar{C}_q^{ii} + \alpha_2 \bar{R}_{ij,q}^{ij}\right) \leq x_q, \qquad q = 1,...,n \; (I)$$
$$\alpha_1 + \alpha_2 = \frac{max\{\bar{p}_q\}}{max\{\bar{C}_q^{ii}\}}, \qquad q = 1,...,n \; (II)$$
Variables: $\alpha_1, \alpha_2, x_q \geq 0, \quad q = 1,...,n$

$$min \; \sum_{q=1}^{n} y_q \qquad (4)$$
Subject to:
$$\bar{m}_q - \left(\beta_1 \bar{C}_q^{ii} + \beta_2 \bar{R}_{ij,q}^{ij}\right) \leq y_q, \qquad q = 1,...,n \; (I)$$
$$\beta_1 + \beta_2 = \frac{max\{\bar{m}_q\}}{max\{\bar{C}_q^{ii}\}}, \qquad q = 1,...,n \; (II)$$
Variables: $\beta_1, \beta_2, y_q \geq 0, \quad q = 1,...,n$

## V. PERFORMANCE EVALUATION

To assess the performance of the proposed model in Eq. (2), the open-source Asterisk software [20] is used to implement SIP severs. Besides, to implement user agent, the open source SIPp software [21] is used. Finally, to implement LB-CAC and solve the mathematical models, MATLAB software is utilized. All considered servers and LB-CAC in this study have a homogenous hardware composing of an INTEL Dual Core 3GHZ CPU and a 512 MB memory, using a Linux CentOS v.6.3 operating system. Moreover, the reports of Asterisk software are used to measure call status, and Oprofile software is utilized to measure CPU and memory usages. To compute coefficients $\alpha$ and $\beta$, some (random) local and outbound calls are established in the server and CPU and memory usages of the server are measured. This experiment is repeated for 100 times ($h = 100$) and the obtained data are gathered in a dataset. By solving models Eq. (3) and Eq. (4), $\alpha_1$ and $\alpha_2$ are determined as 0.074104 and 0.025896, respectively; while $\beta_1$ and $\beta_2$ obtaining as 0.327393 and 0.184607, respectively[2].

---

[2] We prepared this dataset by conducting some tests on a real testbed in IP-PBX laboratory. The complete information of the dataset is available at: http://ippbx-lab.um.ac.ir/index.php?&newlang=eng

## A. Simulations and the Numerical Results

To evaluate the performance of the proposed heuristic method, we consider the topology shown in Fig. 5; moreover, the following three scenarios in Table 1 are investigated with different $\gamma$ and $\varphi$ (represented in four cases: *f1* to *f4*). Here, $M_l$ and $P_l$ ($l: 1, ..., n$) are considered as 512 and 100 for all servers, respectively. Also $\tau$ is considered as 3 sec for all runs. Fig. 6 and Fig. 7 illustrate optimal values of $p_l$ and $m_l$ for all servers. In addition, Fig. 8 presents optimal call admission rate for different states.

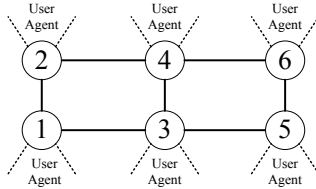

Fig. 5. The studied topology

TABLE I. THE SCENARIOS

| $\mathbb{C}^{ij}$ | | | | | | $\sum_{i=1}^{n}\sum_{j=1}^{n} \mathbb{C}^{ij}$ |
|---|---|---|---|---|---|---|
| 10 | 20 | 4 | 30 | 60 | 8 | |
| 10 | 50 | 20 | 10 | 4 | 10 | |
| 0 | 0 | 100 | 30 | 40 | 12 | |
| 10 | 20 | 30 | 46 | 50 | 14 | 860 |
| 50 | 6 | 30 | 20 | 10 | 20 | |
| 6 | 40 | 25 | 40 | 10 | 15 | |
| Scenario 1 (low load) | | | | | | |
| 100 | 110 | 64 | 65 | 96 | 58 | |
| 60 | 95 | 70 | 70 | 92 | 90 | |
| 40 | 65 | 120 | 80 | 110 | 92 | |
| 50 | 80 | 70 | 86 | 110 | 94 | 2853 |
| 90 | 76 | 60 | 50 | 70 | 80 | |
| 46 | 95 | 70 | 94 | 70 | 85 | |
| Scenario 2 (medium load) | | | | | | |
| 110 | 120 | 84 | 80 | 105 | 65 | |
| 70 | 105 | 80 | 75 | 100 | 98 | |
| 50 | 75 | 125 | 90 | 120 | 98 | |
| 60 | 90 | 80 | 95 | 115 | 100 | 3188 |
| 100 | 86 | 80 | 60 | 74 | 80 | |
| 66 | 105 | 80 | 104 | 78 | 85 | |
| Scenario 3 (high load) | | | | | | |

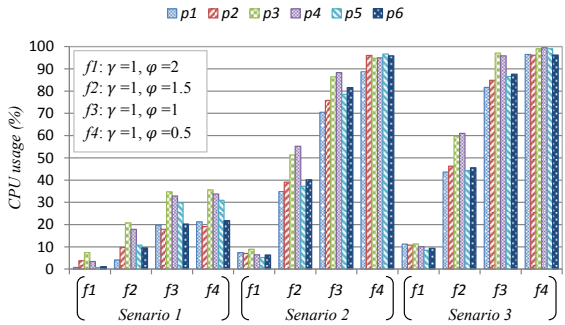

Fig. 6. The optimal CPU usages ($p_l$)

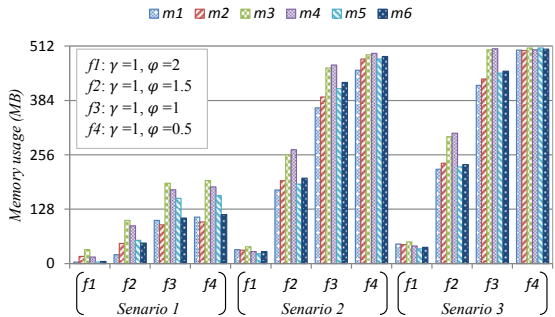

Fig. 7. The optimal memory usage ($m_l$)

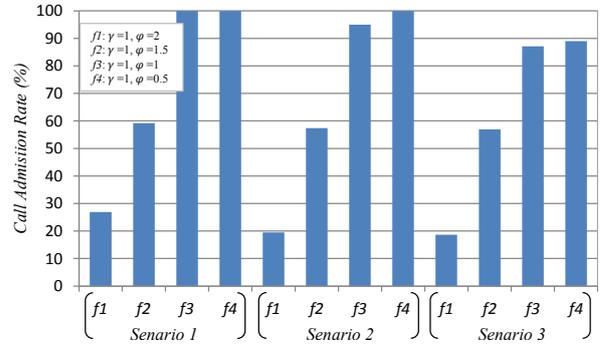

Fig. 8. The optimal call admission rate

In Fig. 6, Fig. 7, and Fig. 8, for all scenarios, by considering *f1* to *f4*, the resource usage and call admission rate show increasing trend. In this regard, the ratio between parameters $\gamma$ and $\varphi$ illustrates the importance of call admission or resource preservation. In case *f1*, resource preservation is more significant as compared to case *f4* and even results in blockage of some calls. In contrast, in case *f4*, the maximum admission of the calls is preferred, even if it results in higher resource consumption (Fig. 8). By making a trade-off between theses parameters, it is possible to promote the call admission rates and optimal usage of the resources.

Moreover, regarding the amount of load from scenario 1 to scenario 3, Fig. 6 and Fig. 7 reveal that resource usages would be raised; however, the load input of the network must be that high that even using the entire resource, it is not possible to respond all input calls (Fig. 8, scenario 3, cases *f3* and *f4*).

In scenario 1, the SIP network does not enter the overload stage, as all input load of the network is admissible by using almost limited amount of resource (Fig. 8, scenario 1, cases *f3* and *f4*). For instance, in this scenario, in case *f3*, optimal $m_1$ and $p_1$ for admitting all input calls are 101.3384 and 19.68201, respectively. In scenario 2, in which the input load is higher than that of scenario 1, maximum call admission rate can be obtained in case *f4* (Fig. 8). In *f4*, however, the resources are almost entirely used (Fig. 6 and Fig.7). The comparison of Fig. 6, Fig. 7, and Fig. 8 indicates that in scenario 3 even using the entire servers' resources, it is not possible to reach optimum call admission rate greater than 89%, as the input load would exceed the network capacity and the extra load would be blocked.

By solving model in Eq. (2), in addition to determination of the optimal values of $m_l$, $p_l$, and $C^{ij}$, the optimal $R_{kl}^{ij}$ is specified. For instance, in scenario 3 and case *f4*, the total call requests from path 1 to 6 ($\mathbb{C}^{16}$) is 65 in which 27.5 is admitted ($C^{16}$). Distribution of these admitted calls is shown in Fig. 9. As it is shown, $C^{16}$ is distributed between two paths to reach the maximum objective function as Path 1: ($R_{13}^{16} = 11.2$, $R_{35}^{16} = 11.2$, $R_{56}^{16} = 11.2$) and path 2: ($R_{12}^{16} = 16.3$, $R_{24}^{16} = 16.3$, $R_{46}^{16} = 16.3$). For all cases *f1* to *f4*, the load is distributed among the servers in a way that maximum objective function is achieved. The average time for each run ($t_c$) is almost 0.95 sec, which can be ignored considering the length of one call.

## B. Implementation and the Experimental Results

To evaluate operational performance of the heuristic method, a testbed with topology shown in Fig. 5 is prepared using Asterisk servers). To this aim, $C^{ij}$ and $R_{kl}^{ij}$ values obtained in the previous part for scenario 2, cases *f2* and *f4* are injected to testbed and then the numbers of successful calls as well as the CPU and memory usages are measured. Note that, if $C^{ij}$ and $R_{kl}^{ij}$ are not integer numbers, they would be rounded down. Fig. 10 and Fig. 11 indicate that the performance of the servers and optimal computations of LB-CAC yield rather similar results. It is found that, due to some extra operations in the testbed's servers, the implementation values are slightly higher than the simulation ones. However, this difference can be minimized by excluding the extra modules e.g. billing module. The links used in the testbed are 1 Gbps. While adding bandwidth constraint into the proposed model in Eq. (2) can be easily performed, while the obtained model can be remained in LP form.

As illustrated in Table 1, the total number of call requests in scenario 2 is 2,853 among which 1,638 requests are admitted in case *f2*, so that an admission rate of 57.41% is obtained (Fig. 8). On the other hand, only 7 calls from 1,638 call requests in the testbed are failed. As illustrated in Table 2, repeating this experiment clarifies that by rounding down the results of analytical values from LB-CAC, it does not seriously influence on the number of successfully serviced calls (measured in testbed) and the constraints of the proposed model in Eq. (2) are still in feasible space as well.

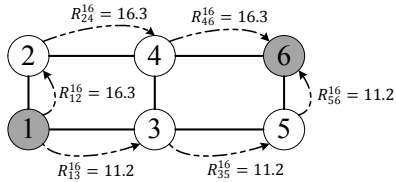

Fig. 9. Load distribution in paths among servers 1 through 6 (signaling paths)

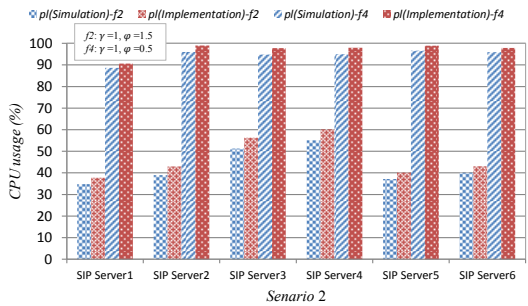

Fig. 10. Comparison between controller $p_l$ and implementation case $p_l$

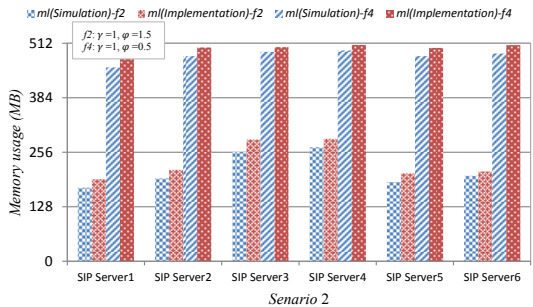

Fig. 11. Comparison between controller $m_l$ and implementation case $m_l$

TABLE II. A COMPARISON BETWEEN THE NUMBER OF ADMITTED AND SERVICED REQUESTS

|  | Scenario 2 | | | | Scenario 3 | | | |
|---|---|---|---|---|---|---|---|---|
|  | *f1* | *f2* | *f3* | *f4* | *f1* | *f2* | *f3* | *f4* |
| Admitted by LB-CAC | 556 | 1638 | 2710 | 2853 | 594 | 1815 | 2777 | 2837 |
| Serviced in testbed | 552 | 1631 | 2704 | 2845 | 589 | 1806 | 2767 | 2828 |

## VI. CONCLUSION AND FUTURE WORK

Overload phenomenon in SIP signalling network occurs when a SIP server does not have enough resources to process messages. We showed that the problem of overload control in SIP network with $n > 2$ servers and limited resources is in the form of NP-hard. Then, we introduced a Load-Balanced Call Admission Controller (LB-CAC) based on a heuristic mathematical model to determine an optimal resource allocation in a way that maximizes the number of requested local and outbound calls. In fact, the linear optimization model was proposed to maximize call admission rate along optimal allocation of CPU and memory resources of the SIP servers. Comparison the analytical and experimental results in various scenarios showed the efficiency of the proposed method. Developing the proposed method in distributed form (instead of a central controller) can be mentioned as a future work.